# On forbidden high-energy electrons as a source of background in X-ray and gamma-ray observations

Alla V. Suvorova[1,2] and Alexei V. Dmitriev[1,2]

[1] Institute of Space Science, National Central University, No 300 Jungda Rd, Jhongli, Taoyuan 32001, Taiwan

[2] Skobeltsyn Institute of Nuclear Physics, Lomonosov Moscow State University, Moscow 119234, Russia

Correspondence should be addressed to Suvorova Alla; alla@jupiter.ss.ncu.edu.tw

**Abstract**

The study is devoted to a problem of electron-induced contaminant to X-ray and gamma-ray astrophysical measurements on board low-orbiting satellites. We analyzed enhancements of electron fluxes in energy range 100 - 300 keV observed at equatorial and low latitudes by a fleet of NOAA/POES low-orbiting satellites over the time period from 2003 to 2005. It was found that 100-300 keV electron fluxes in the forbidden zone below the inner radiation belt enhanced by several orders of magnitude during geomagnetic storms and/or under strong compressions of the magnetosphere. The enhancements are related to high substorm activity and occurred at any local time. Intense fluxes of the energetic electrons in the forbidden zone can be considered as an essential contaminant to X-ray and gamma-ray measurements at low-latitude and low-altitude orbits.

## 1. Introduction

Astrophysical sources of X-ray and γ-ray emission exist throughout the Universe including galaxies, the solar system and, particularly, the solar corona. To distinguish the emission of a cosmic object from background noise, which is higher in most of cases, is a task of high priority in the observational X-ray and γ-ray astronomy. Identification of all individual sources contributed to the background noise allows improving the techniques for background reduction. Thorough studies show that, particularly, in the X-ray and low-energy γ-ray ranges (from ~100 eV to ~300 keV), the background noise is extremely complex in its origin [1-3]. Nevertheless, two major components in the total background should be distinguished in the first turn: one of astrophysical origin, such as cosmic background, and another of local intrinsic instrumental origin, and besides the last affects on measurement accuracy of the first. As it was recently pointed out, the most important effect limiting the accuracy of cosmic X-ray background measurements is related to the intrinsic background variation in detectors [4-6].

Astronomical observations are performed on board high- and low-orbiting spacecrafts or on space platforms, in the Earth upper atmosphere, magnetosphere and interplanetary space. Note that orbit choice is one of the important factors affecting the sensitivity of the instruments operated under conditions of low signal-to-noise ratio. In either case the instruments encounter an intense radiation environment, so the background noise is estimated in a specified radiation environment. The internal instrumental noise is caused by the interactions of energetic charged particles with the spacecraft and detectors. Either by direct penetration or by secondary radiations produced in the payload materials, photon detectors may at times give spurious responses, particularly if the "background" radiations are nonsteady [7].

Vast majority of astronomy satellites are low-orbiting, i.e. they pass through zones of enhanced particle radiations, the Earth's auroral zone and radiation belt (RB) at high latitudes and also South Atlantic Anomaly (SAA) at low latitudes. Knowledge of contaminating background radiation of magnetospheric origin is accumulated since 60s and the influence of RB has been well investigated [1, 8-10]. Particularly, it was shown that detectors of X- and γ-ray are subject to in-orbit enhanced background noise caused by the magnetospheric high-energy electrons (e.g., [11-16]).

To minimize the charge particle contamination from RB, most astronomy experiments are conducted at low inclination orbits, for example RHESII mission [17]. It is considered that the equatorial region away from the SAA is characterised by low background, so measurements made during passages of the spacecraft through the equator-to-low latitudinal region are particularly valuable to high-energy astrophysics. However, even during the early 1970s, the X-ray astronomers revealed unexpected electron-induced contaminant at relatively low latitudes as well, which was a few times higher than the cosmic X-ray background [7, 18]. They found several events when flux intensity of electrons with energy of tens of keV was as large as $\sim 10^3$ el/(cm$^2$ s sr)$^{-1}$, exceeding quiet level by 2 orders of magnitude. Note that it is still much lower than in the radiation belt and auroral zones (see [14]). In addition to astrophysical measurements, ionospheric and atmospheric studies [19-26] and satellite data failures studies (e.g., [27-29]) also found several effects suggesting that electron impact is important factor at low and middle latitudes. That is, more importantly the occasional increases of electron flux below the radiation belt were discovered even earlier in direct satellite-borne measurements [30, 31] and then corroborated in several studies [32-35]. They reported about sporadic fluxes of very high intensity, which was comparable with the auroral precipitation. However, the direct observations of sporadic events caused strong argument due to a doubt about validity of measured high intensity (see review by Paulikas [36]). As a result of this, despite importance of low-latitude measurements of electron enhancements recognized earlier [9] further investigation of this phenomenon was not carried out.

Until now, sporadic enhancement of energetic electrons below the inner radiation belt (IRB) is a poor-studied phenomenon [24, 37]. Comprehensive studies based on large statistics collected for more than ten years [38-42] have showed that fluxes of quasi-trapped electrons within the wide energy range 10-300 keV can increase dramatically by a few orders of magnitude relative to the quiet level at very low L shells ($L < 1.1$), so-called forbidden zone. The most extreme intensity of forbidden zone fluxes of the same order as auroral precipitations, $\sim 10^6 - 10^7$ (cm$^2$ s sr)$^{-1}$, was observed during major magnetic storms driven by a coronal mass ejection. Nevertheless, major storms themselves are not a necessary condition for electron enhancements in the forbidden zone. Another important solar wind driver resulting in significant flux enhancements is the extremely strong solar wind dynamic pressure, as it occurred on 21 January 2005 [42, 43].

In the present work, we focus on analysis of great enhancements of the forbidden electrons in the high-energy range 100-300 keV observed by a fleet of NOAA/POES low-orbiting satellites during the declining phase of the 23$^{rd}$ solar cycle over the time period from 2003 to 2005. Note, that the enhancements within the SAA region are excluded from our consideration. We demonstrate temporal and spatial characteristics of forbidden electron enhancements during different types of magnetic storms.

## 2. Experimental Data

We used data on >100 and >300 keV electron fluxes measured on board the polar orbiting NOAA/POES satellites [44]. The POES satellites have Sun-synchronous orbits at altitudes of ~ 800 - 850 km.

It is well known that the electron measurements can be distorted by proton contamination and nonideal detector efficiency. According to a comprehensive study [45], the 100 keV electron fluxes practically do not change, while the 300 keV electron fluxes should be decreased by about twenty percent. Because this factor is not crucial for the current study, we present uncorrected fluxes.

## 3. Enhancements of High-Energy Electrons

Figure 1 presents geographical distributions of high-energy electrons in two energy ranges >100 keV (on the left) and >300 keV (on the right) during three major geomagnetic storms on 29-31 October 2003, 7-10 November and 15-16 May 2005. Each global map was compiled from measurements of two orthogonally oriented detectors (0°-detector and 90°-detector) of three POES satellites (NOAA-15, NOAA-16, NOAA-17). In each spatial bin, the maximal value of electron flux instead of the averaged one is shown.

In Figure 1, one can easily distinguish typical regions of the stably trapped radiation: the outer radiation belt and auroral precipitation zone at high latitudes (>60°), the SAA area at low latitudes with a core in vicinity of -60° longitude. In addition at low latitudes outside SAA, it is also clearly seen intense electron fluxes, which were occasionally observed at some longitudes. Note that this region is situated below IRB at L-shells less than ~1.1. In the literature, this region is called a forbidden zone, which is characterized by tenuous fluxes of ~1-10 $(cm^2 \, s \, sr)^{-1}$. Nevertheless, in three cases presented, the largest intensity of the 100 keV electrons in the forbidden zone exceeds $10^6$ $(cm^2 \, s \, sr)^{-1}$.

In [41] it is shown that forbidden energetic electrons are quasi-trapped particles with pitch angles range within the drift loss cone (i.e. locally trapped). These electrons cannot close the full drift shell around the Earth because they descent dramatically in the SAA area and eventually they are lost in the atmosphere. Therefore, it is reasonably to expect that duration of an enhanced flux phenomenon should be less than a drift period of electrons. At heights of ~900 km, 100 keV and 300 keV electrons have the drift period of ~6 and ~2 hours, respectively (see [46]). In this connection, it is interesting to note that the great enhancements in Figure 1 were observed at longitudes, which are widely separated from each other by ~180°. That indicates to a large scale and/or long duration of the phenomenon. As for higher energy range, >300 keV, one can see a few burst-like enhancements of moderate intensity $10^4 - 10^6$ $(cm^2 \; s \; sr)^{-1}$.

Using ten-year data of the POES's measurements, we found that durations of the enhancements varied from an hour to ten hours and also can be observed at any local time. We analyze time variations of 100 keV electron fluxes in the forbidden zone by considering the following types of geomagnetic disturbances: a major storm, moderate storm and strong magnetospheric compression by the solar wind plasma.

Figure 2 presents time profiles of 100 keV electron fluxes observed by POES satellites from 4 to 18 UT during the major storm on 15 May 2005 (see bottom panel in Figure 1). The electron flux below IRB (black curves) at some longitudes exceeded the quiet level by ~ 4 – 6 orders of magnitude. In Figure 2, one can see that sometimes the intensities in the forbidden zone approached an extremely large value of $10^7$ $(cm^2 \; s \; sr)^{-1}$, for example, after 7 UT (P5) and around 11 UT (P6, P7). The maximal intensity is the same as in the IRB (the SAA area), for example, see P5 around ~11 UT and P7 around ~14 UT, and even higher than in the outer radiation belt or auroral zone (gray curves in Figure 2).

The duration of 100 keV electron enhancements is one of most prolonged (about ten hours) beginning at ~7 UT and ending at ~16 UT. In Figure 2, geomagnetic activity is presented by a storm ring current (*Dst*-variation) and auroral electrojet index (*AE*), which both change in different manner. The *Dst* value varies smoothly, while the *AE*-index shows successive intensifications corresponding to substorm activations. Initially, the electron enhancements appeared during the maximum of geomagnetic activity revealed in both strong negative *Dst*-variation (~-300 nT) and very high AE index (~1800 nT). The geomagnetic activity, both in *Dst* and *AE* indices, was decreasing after 10 UT. But the electron enhancements were observed continuously during the next several hours on the storm recovery phase.

Another important characteristic, which we want to point out, is that the enhancements can be observed at any local time (LT), including both nighttime and sunlit side of the near-Earth orbit. Notably, the P5 satellite observed events in the early morning (6 LT) and evening (18 LT), respectively, at ~7 UT, 14 UT, 16 UT and 8 UT, 10 UT; the P6 satellite – at night (3 LT) at ~11 UT and ~1230 UT; the P7 satellite – at daytime (10 LT) and night (22 LT), respectively, at ~7 UT, 0830 UT and 0930 UT, 11 UT.

As for higher energy electrons of >300 keV, the flux increased at nighttime (2 LT and 22 LT) around 11 UT and at daytime (10 LT) around 7 UT as well (not shown). The duration of 300 keV electron enhancements was shorter than of 100 keV electrons (about of 4 hours).

In Figure 3, we present a moderate storm event on 7-8 January 2005. We distinguish two separated time intervals, during which the 100 keV enhancements were observed: from 16 to 18 UT on 7 January and from 0 UT to 5 UT on 8 January (24 UT and 29 UT, respectively, in Figure 3). Flux intensities during the first interval were more than one order smaller than during the second one. The first enhancements (see P5 and P7) were observed at local morning and prenoon (6 LT and 10 LT). Enhancements during the second interval (see P5 and P6) were observed again at local morning and afternoon (6 LT and 15 LT). Recurring enhancements at local morning can be considered as abundant evidence that both events occurred independently, because, as mentioned above, the drift period of electrons is relatively short. Note that in this event, >300 keV electrons increased only once at dusk (18 LT) closely to a vicinity of SAA (not shown). Note that a substorm (AE increase) occurred before each electron enhancement event.

Figure 4 shows 100 keV electron enhancements during the prolonged compression of the magnetosphere by extremely high solar wind dynamic pressure of more than 150 nPa [43]. Note that the magnetic storm was of moderate strength. Due to the compression, the Earth's magnetopause shrunk to about ~3-4 Re in the subsolar region, radiation belt and ring current moved closely to the Earth. The enhancements were observed at ~19 UT by P5, at 2030 UT by P7 and at 22 UT by P6, with all three occurring during daytime (6 LT, 10 LT and 15 LT). Thus, this case is a good example of single event with a short duration of about 3 hours, when the intense fluxes of 100 keV electrons disappeared within approximately three hours due to their relatively fast azimuthal drift. In this event, >300 keV electrons were not observed at all. Note also, that there was only one (isolated) substorm (AE-index), which is seems relating to this short event.

## 4. Discussion

We have demonstrated several examples of unusual latitudinal distributions of high-energy electrons during geomagnetic disturbances such as strong magnetic storms and magnetospheric compression. Figure 1 demonstrates that electron fluxes at equator-to-low latitudes extremely increase and sometimes extended almost globally. This picture doesn't conform to a general view that is a significant population of permanently trapped electrons can not present within the low-latitudinal region not only during quiet time but also during storms. How often do high-energy electrons enhance in the forbidden zone and why?

Figure 5 shows global maps of >100 keV and >300-kev electrons compiled from measurements of the detector pointed to zenith for three years 2003, 2004, and 2005 during the declining phase of the 23$^{rd}$ solar cycle. It is clearly seen from comparison of Figure 5 with Figure 1 that contributions of the presented cases to the corresponding one-year distributions are the biggest, probably because these major storms were either of super-storm series (October 2003 and November 2004) or isolated but the strongest one (May 2005). Also, note that enhancements in both energy range are moderate ~$10^4 - 10^5$ (cm$^2$ s sr)$^{-1}$ during the rest of geomagnetic disturbances.

It is hard to predict the forbidden electron enhancements, because a mechanism of the process is not established yet. First of all, one should determine the source of forbidden electrons. As discussed in paper [47], one of most likely candidates is the radiation belt. But the process of particle penetration from the radiation belt to the forbidden zone is still unclear. It might be a fast radial drift in strong electric fields of ~20 mV/m occurred at L-shells below 2. However, there is a lack of experimental measurements in this region. Hence, this concern needs further experimental data and modeling.

Meanwhile, one can expect that if energetic electrons occasionally enhanced in the forbidden zone then duration of the event could be comparable with or even less than the period of azimuthal drift of an energetic electron around the Earth. For example, 100 keV electrons, having the drift period of ~6 hours at a height of ~900 km, can pass from Japan to SAA (~120° in longitudes) for two or three hours (see [46]). While 300 keV electrons drift eastward approximately 3-4 times faster and spend ~30 minutes for running ~120° separation in longitudes. Hence, a sporadic enhancement, resulted from a single short-time electron injection, can be observed by several satellites at different locations but within a relatively short period of time. An example of such behavior is the case on 21-22 January 2005 shown in Figure 4 (see also [42]). After the substorm expansion, the forbidden electrons were observed by different POES satellites in the longitudinal range from 160°E to 110°W (~90° longitude interval) during three hours.

However, from observations of 100 keV electrons, we found many long-lasting enhancements (see Figures 2 and 3). Two events in the case of 7-8 January 2005 are separated by 8-hours interval. Taking into account the drift period of ~6 hours, we can conclude that those enhancements occurred independently and they are probably driven by two successive substorms revealed in the increases of AE. Similar causal relation between the substorms and enhancements can be found in the case of 15 May 2005. Both cases support our suggestion that strong substorm serves as indicator of a process resulting in forbidden electron enhancements.

We can also suggest that the long-lasting high-energy electron enhancements relate to the long-lasting high auroral activity. We conclude that in these cases a prolonged or multiple triggering resulted in long-lasting high-energy electron enhancements.

Thus, the forbidden electron enhancements relate closely to a geomagnetic activity and, hence, are driven by the variations in solar wind dynamic pressure and interplanetary magnetic field [41, 42, 47].

Finally, it is important to point out that forbidden electron enhancements in a higher energy range (>300 keV) look mostly like burst events (see Figure 1, right panels). Moreover, they can be observed at any local time, both nighttime and sunlit side of the near-Earth orbit. Such feature makes it difficult to recognize the astrophysical X-ray source (solar, cosmic or galactic) from the high-energy electron-induced events. The forbidden high-energy electrons will increase detector noise, which may mask the original source signal.

## 5. Summary

In the paper we consider a very important problem of the electron contamination to high-energy astrophysical and solar hard radiation measurements. The study is based on long-term statistics of the high-energy electron observations by low-orbiting satellites. We have demonstrated that 100-300 keV electron fluxes significantly exceeded a quite level during major geomagnetic storms and strong compressions of the magnetosphere as well. Enhanced fluxes of the forbidden high-energy electrons can significantly contaminate to X-ray measurements at low-latitude and low-altitude orbit.

**Conflict of Interests**

The authors declare that there is no conflict of interests regarding the publication of this paper.

**Acknowledgements**

The authors thank a team of NOAA's Polar Orbiting Environmental Satellites for providing experimental data about energetic particles. The work of Alla V. Suvorova was supported by grant NSC-102-2811-M-008-045 from the National Science Council of Taiwan. Alla V. Suvorova and Alexei V. Dmitriev gratefully acknowledge the support of part of this work by grant NSC103-2923-M-006-002-MY3/14-05-92002HHC_a of Taiwan-Russia Research Cooperation

**References**

[1] A. J. Dean, F. Lei, and P. J. Knight, "Background in space-borne low-energy γ-ray telescopes", *Space Sci. Rev.*, vol. 57, no. 1, pp. 109-186, 1991.

[2] R. Smith, S. Snowden, K. D. Kuntz (Eds.), "The Local Bubble and Beyond II", Philadelphia, USA, 21-24 April, 2008, *AIP Conference Proceedings/Astronomy and Astrophysics,* vol. 1156, 2009.

[3] T. Takahashi, S.S. Murray, J.-W. A. den Herder (Eds.), "Space Telescopes and Instrumentation 2012: Ultraviolet to Gamma Ray", Amsterdam, Netherlands, 1-6 July, 2012, *Proceedings of SPIE* vol. 8443, SPIE, Bellingham, WA, doi: 10.1117/12.2008904, 2012.

[4] D. E. Gruber, J. L. Matteson, L. E. Peterson, and G. V. Jung, "The spectrum of diffuse cosmic hard X-rays measured with HEAO 1", *The Astrophys. J.*, vol. 520, pp. 124-129, 1999.

[5] G. S. Bisnovatyi-Kogan and A.S. Pozanenko, "About the measurements of the hard X-ray background", *Astrophys. Space Sci.*, vol. 332, pp. 57–63, doi: 10.1007/s10509-010-0485-9, 2011.

[6] M. Revnivtsev et al., "MVN: x-ray monitor of the sky on Russian segment of ISS", in Space Telescopes and Instrumentation 2012: Ultraviolet to Gamma Ray, edited by T. Takahashi, S.S. Murray, J.-W. A. den Herder, Amsterdam, Netherlands, 1-6 July, 2012, *Proceedings of SPIE* vol. 8443, SPIE, Bellingham, WA, doi: 10.1117/12.925916, 2012.

[7] F. D. Seward, R. J. Grader, A. Toor, G. A. Burginyon, and R. W. Hill, "Electrons at low altitudes: A difficult background problem for soft X-ray astronomy", *Proceedings of the Workshop on Electron Contamination in X-Ray Astronomy Experiments*, edited by S.S. Holt, NASA GSFC Rep. X-661-74-130, 1974.

[8] J. B. Cladis and A. J. Dessler, "X rays from Van Allen belt electrons", *J. Geophys. Res.*, vol. 66, no. 2, pp. 343-350, 1961.

[9] S. S. Holt, (Ed.), *Proceedings of the Workshop on Electron Contamination in X-Ray Astronomy Experiments*, Washington, 26 April 1974, NASA GSFC Rep. X-661-74-130, 1974.

[10] K. S. K. Lum, J. J. Mohr, D. Barret, et al., "Simulations and measurements of the background encountered by a high-altitude balloon-borne experiment for hard X-ray astronomy", *Nuclear Instruments and Methods in Physics Research, Section A*, vol. 396, pp.350-359, 1997.

[11] W. C. Feldman, E.M.D. Symbalisty, and R.A. Roussel-Dupre, " Hard X-ray survey of energetic electrons from low-Earth orbit", *J. Geophys. Res.*, vol. 101, no. A3, pp. 5195–5209, 1996.

[12] U. B. Jayanthi, M. G. Pereira, I. M. Martin, et al.,"Electron precipitation associated with geomagnetic activity: Balloon observation of X ray flux in South Atlantic anomaly", *J. Geophys. Res.*, vol. 102, no. A11, pp. 24069-24073, 1997.

[13] R. Bucik, K. Kudela, A. V. Dmitriev, S. N. Kuznetsov, I. N. Myagkova, and S. P. Ryumin, "Spatial distribution of low energy gamma-rays associated with trapped particles", *Adv. Space Res.*, vol. 30, no. 12, pp. 2843-2848, 2002.

[14] R. Bucik, K. Kudela, A. V. Dmitriev, S. N. Kuznetsov, I. N. Myagkova, and S. P. Ryumin, "Review of electron fluxes within the local drift loss cone: Measurements on CORONAS-I", *Adv. Space Res.*, vol. 36, no. 10, pp. 1979-1983, 2005.

[15] R. Sarcar, S. Mandal, D. Debnath, et al., "Instruments of RT-2 experiment onboard CORONAS-PHOTON and their test and evaluation IV: background simulations using GEANT-4 toolkit", *Exp. Astron.*, vol. 89, pp. 85-107, doi:10.1007/s10686-010-9208-z, 2011.

[16] A. Gusev, I. Martin, and G.Pugacheva, "The soft X-ray emission of nocturnal atmosphere during the descending phase of the 23rd solar cycle", *Sun and Geosphere*, vol. 7, no.2, pp. 127-131, 2012.

[17] R. P. Lin, B. R. Dennis, G. J. Hurford, et al., "The Reuven Ramaty High-Energy Solar Spectroscopic Imager (RHESSI)", *Solar Physics*, vol. 210, pp. 3-32, 2002.

[18] J. E. Neighbours and G. W. Clark, "A survey of trapped low energy electrons near the inner boundary of the inner radiation zone from the OSO-7", *Proceedings of the Workshop on Electron Contamination in X-Ray Astronomy Experiments*, edited by S.S. Holt, NASA GSFC Rep. X-661-74-130, 1974.

[19] L. A. Antonova and G. S. Ivanov-Kholodny, "Corpuscular hypothesis for the ionization of the night ionosphere", *Geomagn. Aeron.*, vol. 1, no. 2, pp. 164-173, 1961.

[20] L. A. Antonova and T. V. Kazarchevskaya, "Measurements of soft electron streams in the upper atmosphere at altitude to 500 km", *Space Research, Transactions of the All-Union Conference on Space Physics*, ed. by G. A. Skuridin et al., Science Publishing House, Moscow, June 10-16, 1965.

[21] W. C. Knudsen and G. W. Sharp, "F2-region electron concentration enhancement from inner radiation belt particles", *J. Geophys. Res.*, vol. 73, no. 19, pp. 6275-6283, 1968.

[22] R. H. Doherty, "Observations suggesting particle precipitation at latitudes below 40N, *Radio Science*, vol. 6, no. 6, pp. 639-646, 1971.

[23] T. A. Potemra and T. J. Rosenberg, "VLF propagation disturbances and electron precipitation at mid-latitudes", *J. Geophys. Res.*, vol. 78, no. 10, pp. 1572-1580, 1973.

[24] A. V. Dmitriev and H.-C. Yeh, "Storm-time ionization enhancements at the topside low-latitude ionosphere", *Annales Geophysicae*, vol. 26, pp. 867-876, 2008.

[25] A. V. Dmitriev, H.-C. Yeh, M. I. Panasyuk, V. I. Galkin, G. K. Garipov, B. A. Khrenov, P. A. Klimov, L. L. Lazutin, I. N. Myagkova, S. I. Svertilov "Latitudinal profile of UV nightglow and electron precipitations", *Planetary and Space Science*, vol. 59, pp. 733-740, 10.1016/j.pss.2011.02.010, 2011.

[26] P. Bobik, M. Putis, M. Bertaina, et al., "UV night background estimation in South Atlantic Anomaly", *Proceedings of the 33rd International Cosmic Ray Conference (ICRC)*, Rio de Janeiro, Brazil, 2-9 July 2013.

[27] A. V. Dmitriev, I. I. Guilfanov, M. I. Panasyuk, "Data failures in the "Riabina-2" experiment on MIR orbital station", *Rad. Meas.*, vol. 35, no. 5, pp. 499-504, 2002.

[28] J. F. Fennell, J.L. Roeder, and H.C. Koons (2002), "Substorms and magnetic storms from the satellite charging perspective", *COSPAR Colloquia Series*, 12, 163-173, 2002.

[29] K. Kudela, "On cosmic rays and space weather in the vicinity of Earth", in *Homage to the Discovery of Cosmic Rays*, edited by J.A. Perez-Peraza, pp. 177-199, NOVA Sci., Inc., New York, 2013.

[30] V. I. Krasovskii, Yu. M. Kushnir, G. A. Bordovskii, G. F. Zakharov, and E. M. Svetlitskii, The observation of corpuscles by means of the third artificial Earth satellite, *Iskusstvennye Sputniki Zemli*, vol. 2, pp. 59-60, 1958 (English translation: *Planet. Space Sci.*, vol. 5, pp. 248-249, 1961).

[31] I. A. Savenko, P. I. Shavrin, and N. F. Pisarenko, "Soft particle radiation at an altitude of 320 km in the latitudes near the equator" (in russian), *Iskusstvennye Sputniki Zemli*, vol. 13, pp. 75–80, 1962 (English translation: *Planet. Space Sci.*, vol. 11, pp. 431-436, 1963).

[32] W. J. Heikkila, "Soft particle fluxes near the equator", *J. Geophys. Res.*, vol. 76, pp. 1076–1078, 1971.

[33] J. D. Winningham, "Low energy (10 eV to 10 keV) equatorial particle fluxes", *Proceedings of the Workshop on Electron Contamination in X-Ray Astronomy Experiments*, edited by S.S. Holt, NASA GSFC Rep. X-661-74-130, 1974.

[34] R. A. Goldberg, "Rocket observation of soft energetic particles at the magnetic equator", *J. Geophys. Res.*, vol. 79, pp. 5299–5303, 1974.

[35] R. Lieu, J. Watermann, K. Wilhelm, J. J. Quenby, and W. I. Axford, "Observations of low-latitude electron precipitation", *J. Geophys. Res.*, vol. 93, pp. 4131-4133, 1988.

[36] G. A. Paulikas, "Precipitation of particles at low and middle latitudes", *Rev. Geophys. Space Phys.*, vol. 13, no. 5, pp. 709-734, 1975.

[37] A. V. Suvorova, L.-C., Tsai, and A. V. Dmitriev, "On relation between mid-latitude ionospheric ionization and quasi-trapped energetic electrons during 15 December 2006 magnetic storm", *Planet. Space Sci.*, vol. 60, pp. 363-369, doi:10.1016/j.pss.2011.11.001, 2012.

[38] A. V. Suvorova, L.-C. Tsai, and A. V. Dmitriev, "On magnetospheric source for positive ionospheric storms", *Sun and Geosphere*, vol. 7, no.2, pp. 91-96, 2012.

[39] A. V. Suvorova, L.-C. Tsai and A. V. Dmitriev, "TEC enhancement due to energetic Electrons above Taiwan and the West Pacific", *Terr. Atmos. Ocean. Sci.*, vol. 24, no. 2, pp. 213-224, doi: 10.3319/TAO.2012.09.26.01, 2013.

[40] A. V. Suvorova, A. V. Dmitriev, and L.-C. Tsai, "Evidence for near-equatorial deposition by energetic electrons in the ionospheric F-layer", *Proc. of International conference "Modern Engineering and Technologies of the Future"*, Krasnoyarsk, Russia, 6-8 Feb. 2013, edited by A. Khnykin, pp. 68-82, Siberian Federal Univ., Krasnoyarsk, Russia,2013.

[41] A. V. Suvorova, A. V. Dmitriev, L.-C. Tsai, V. E. Kunitsyn, E. S. Andreeva, I. A. Nesterov, and L. L. Lazutin, "TEC evidence for near-equatorial energy deposition by 30 keV electrons in the topside ionosphere", *J. Geophys. Res. Space Physics*, vol. 118, pp. 4672–4695, doi:10.1002/jgra.50439, 2013.

[42] A. V. Suvorova, A. V. Dmitriev, and C.-M. Huang, "Energetic electrons below radiation belt contaminating X-ray measurements at low-orbiting satellites", *Journal of Astrophysics*, vol.2014, Article ID 701498, 5 pages, doi:10.1155/2014/701498, 2014.

[43] A. V. Dmitriev, A. V. Suvorova, J.-K. Chao, C. B. Wang, L. Rastaetter, M. I. Panasyuk, L. L. Lazutin, A. S. Kovtyukh, I. S. Veselovsky, and I. N. Myagkova, "Anomalous dynamics of the extremely compressed magnetosphere during 21 January 2005 magnetic storm", *J. Geophys. Res. Space Physics*, vol. 119, pp. 877 – 896, doi:10.1002/ 2013JA019534, 2014.

[44] D. S. Evans and M. S. Greer, "Polar Orbiting Environmental Satellite Space Environment Monitor: 2. Instrument descriptions and archive data documentation", *Tech. Memo. version 1.4*, NOAA Space Environ. Lab., Boulder, Colo, 2004.

[45] T. Asikainen, and K. Mursula, "Correcting the NOAA/MEPED energetic electron fluxes for detector efficiency and proton contamination", *J. Geophys. Res.*, vol. 118, doi:10.1002/jgra.50584, 2013.

[46] L. R. Lyons and D. J. Williams, *Quantitative Aspects of Magnetospheric Physics*, p.231, D. Reidel Pub. Co., Dordrecht Boston, 1984.

[47] A.V. Suvorova, C.-M. Huang, H. Matsumoto, A. V. Dmitriev, et al., "Low-latitude ionospheric effects of energetic electrons during a recurrent magnetic storm", *J. Geophys. Res. Space Physics*, vol. 119, doi:10.1002/2014JA020349, 2014.

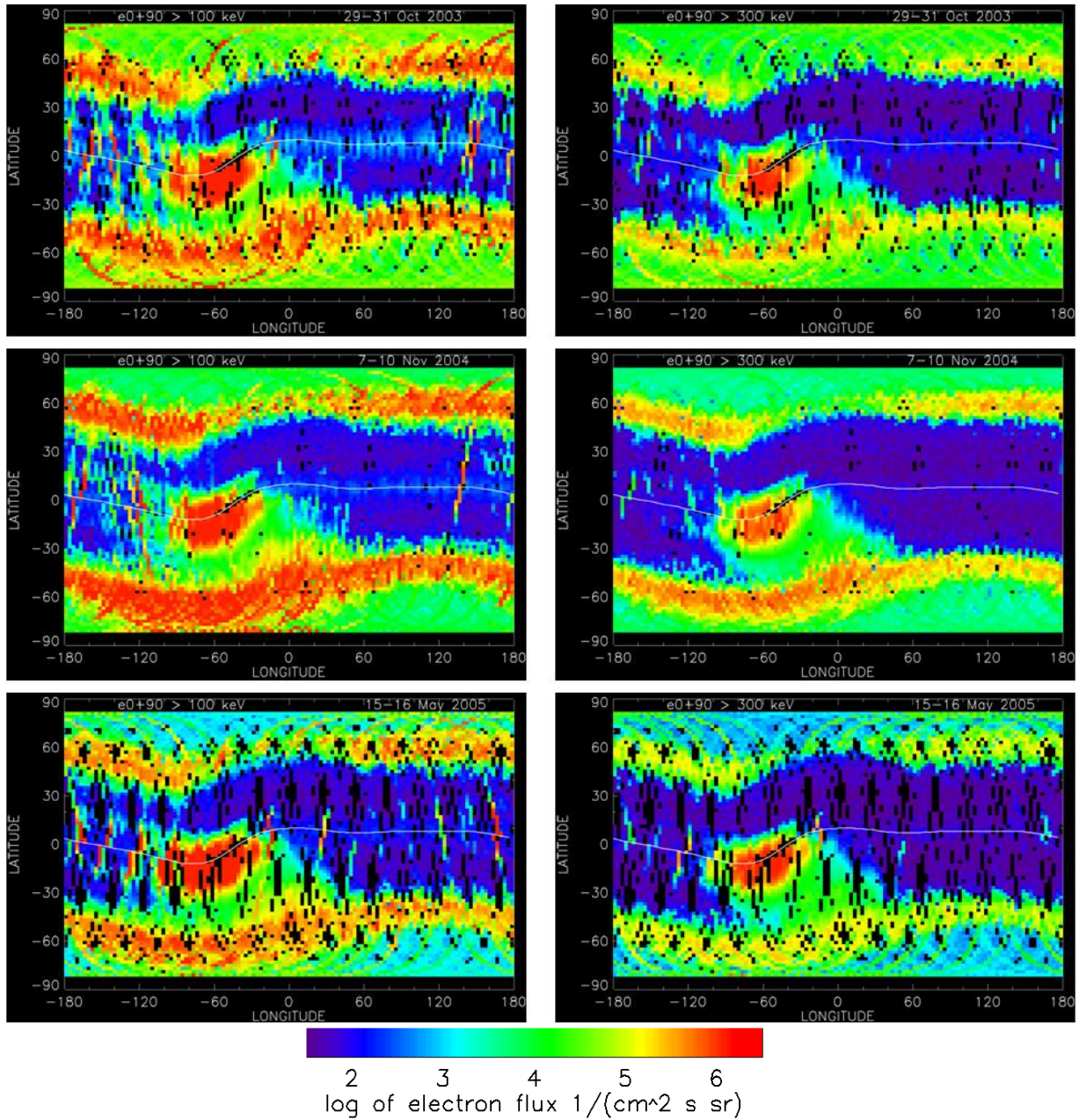

**Figure 1.** High-energy electron enhancements during major geomagnetic storms: (from top to bottom) 29-31 October 2003, 7-10 November 2004, 15-16 May 2005. Geographical maps of global distribution of the electrons in energy ranges >100 keV (left) and >300 keV (right). The maps are composed of data provided by NOAA/POES satellites orbiting at altitude of ~850 km (see details in the text). The white curve indicates the geomagnetic equator. Intensity of high-energy electron fluxes enhances extremely and globally at equator-to-low latitudes (below the inner radiation belt and outside SAA) and even exceeds that one at high latitudes (outer radiation belt and auroral precipitation zone).

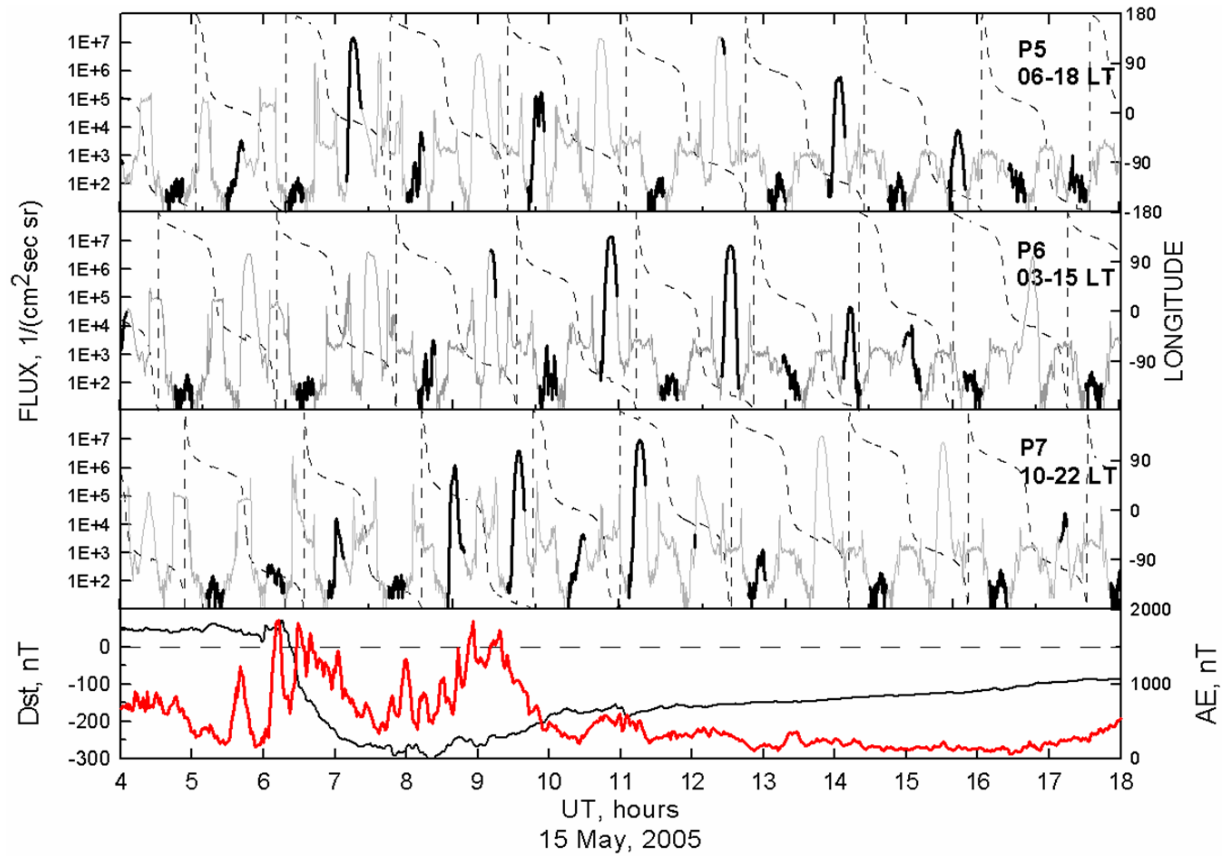

**Figure 2.** Time profiles of >100 keV electron fluxes observed on 15 May 2005 (from top to bottom) by NOAA-15 (P5), NOAA-16 (P6) and NOAA-17 (P7). The satellite local times are indicated on the right. Dashed curves indicate geographic longitudes along the satellite orbit. Black curves correspond to orbital passes at low-latitudes (± 25°) and at all longitudes except the SAA area (approximately between -100° and -20°). Several electron enhancements at low latitudes are seen from 7 UT to 16 UT. The second panel shows geomagnetic activity: the *Dst*-variation (SYM-H index, black curve) and the auroral electrojet (*AE* index, red curve).

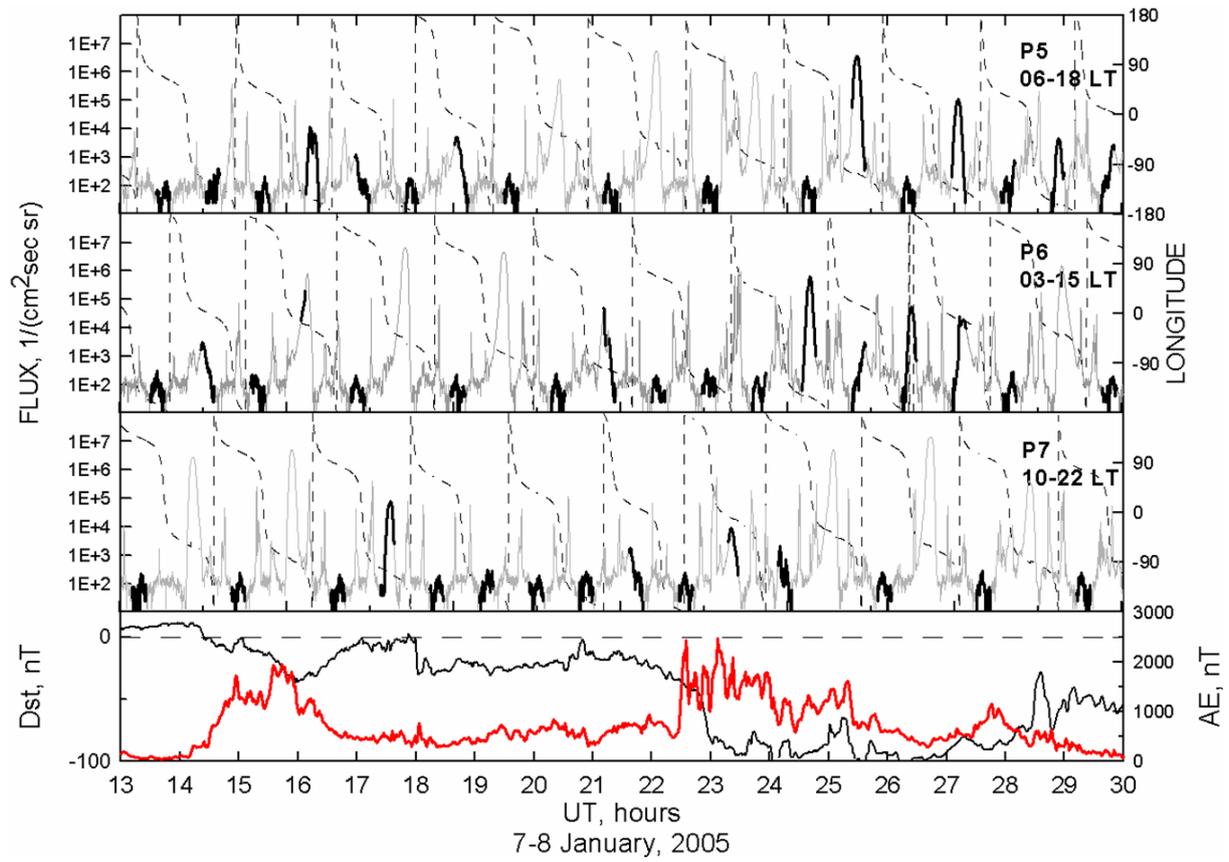

**Figure 3.** The same as Figure 2 but for the moderate storm on 7-8 January 2005.

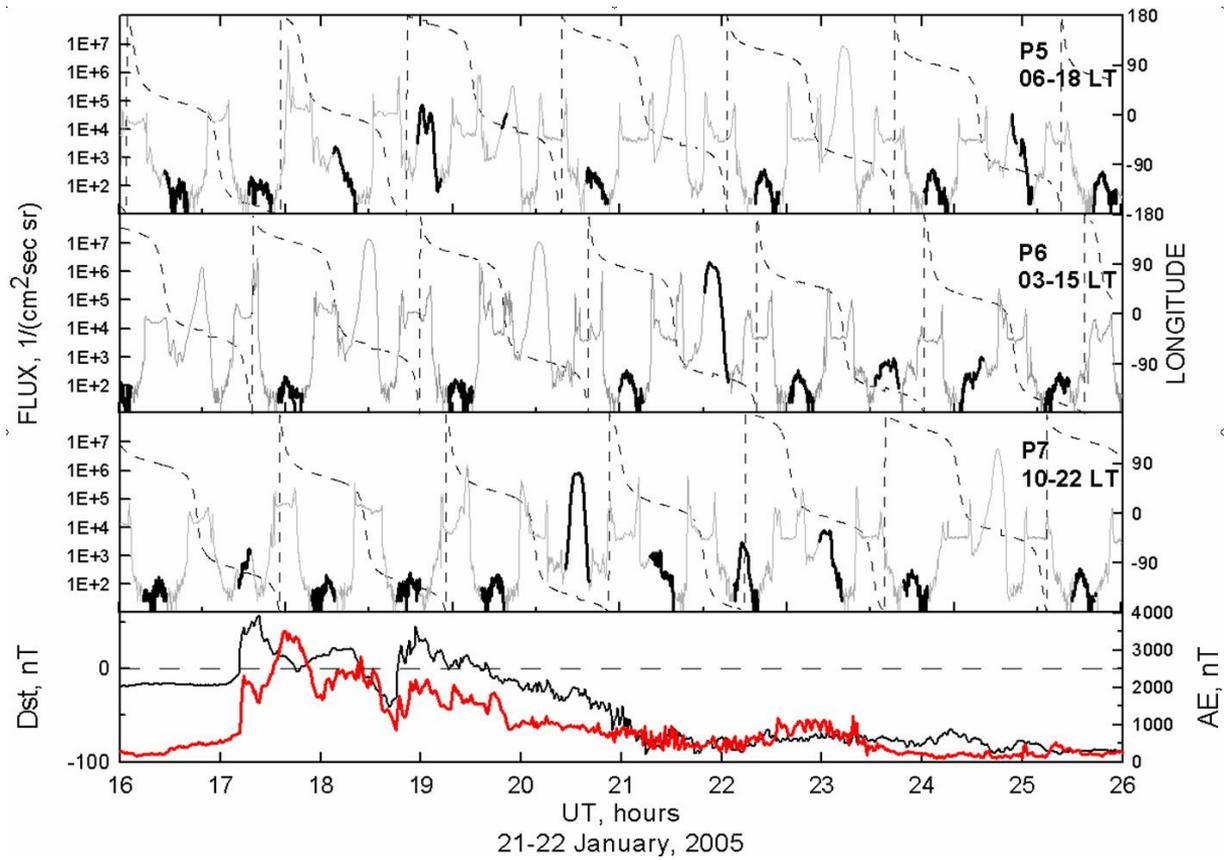

**Figure 4.** The same as Figure 2 but for extremely strong compression of the magnetosphere during a moderate storm on 21-22 January 2005.

E > 100 keV   E > 300 keV

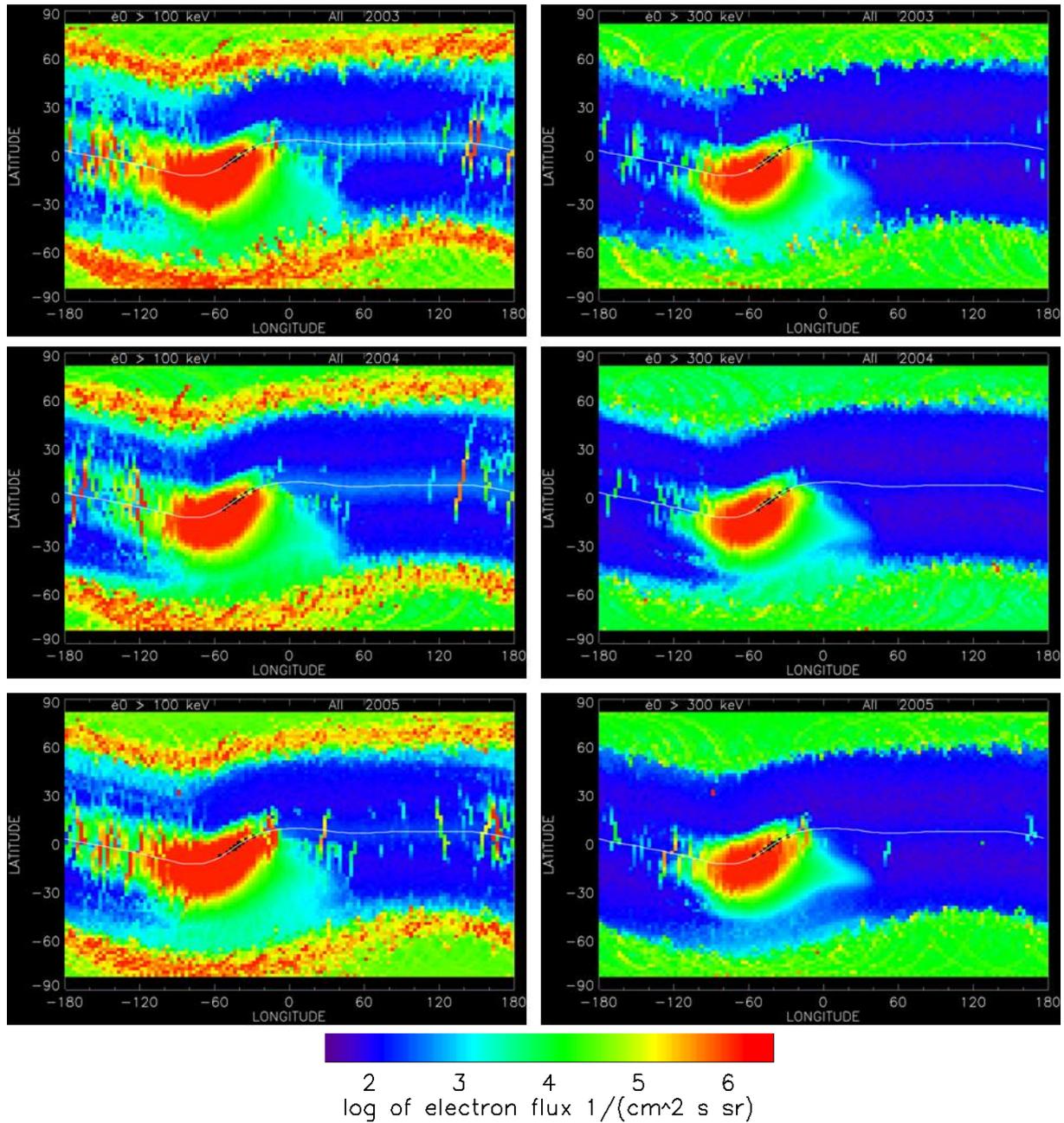

**Figure 5.** The same as Figure 1 but for one-year statistics in: (from top to bottom) 2003, 2004 and 2005 years. Measurements from only the detector pointed to zenith are used (see text for detail).